\documentclass[twocolumn,showpacs,preprintnumbers,amsmath,amssymb,floatfix]{revtex4}
\usepackage{graphicx}
\usepackage{epsfig}

\begin{document}

\title{Large atom number Bose-Einstein condensate of sodium}

\author{K.M.R.~van der Stam}
\affiliation{Atom Optics and Ultrafast Dynamics, Utrecht University,\\
P.O. Box 80,000, 3508 TA Utrecht, The Netherlands}
\author{E.D.~van Ooijen\footnote{Present address: Research Group Ultra Cold Atoms, Niels Bohr Institute,
Universitetsparken 5, 2100 Copenhagen, Denmark}}
\author{R.~Meppelink}
\affiliation{Atom Optics and Ultrafast Dynamics, Utrecht University,\\
P.O. Box 80,000, 3508 TA Utrecht, The Netherlands}
\author{J.M.~Vogels}
\affiliation{Atom Optics and Ultrafast Dynamics, Utrecht University,\\
P.O. Box 80,000, 3508 TA Utrecht, The Netherlands}
\author{P. van der Straten}
\affiliation{Atom Optics and Ultrafast Dynamics, Utrecht University,\\
P.O. Box 80,000, 3508 TA Utrecht, The Netherlands}

\date{\today }

\begin{abstract}

We describe the setup to create a large Bose-Einstein condensate containing more than 120$\cdot$10$^{6}$ atoms. In the experiment a thermal beam is slowed by a Zeeman slower and captured in a dark-spot magneto-optical trap (MOT). A typical dark-spot MOT in our experiments contains 2.0$\cdot$$10^{10}$ atoms with a temperature of $320$ $\mu$K and a density of about 1.0$\cdot$10$^{11}$ atoms/cm$^{3}$. The sample is spin polarized in a high magnetic field, before the atoms are loaded in the magnetic trap. Spin polarizing in a high magnetic field results in an increase in the transfer efficiency by a factor of 2 compared to experiments without spin polarizing. In the magnetic trap the cloud is cooled to degeneracy in 50~s by evaporative cooling. To suppress the 3-body losses at the end of the evaporation the magnetic trap is decompressed in the axial direction.
\end{abstract}
\pacs{32.80.Pj}

\maketitle

\section{Introduction}

Since the first realization of a Bose-Einstein condensate (BEC) 10 years ago \cite{Rb,Na,Li}, almost 60 groups built a setup, in which this phase transition is achieved. The setups lead normally to BEC's containing 1.6$\cdot$10$^{3}$ \cite{Hansch} up to 3$\cdot$10$^{7}$ \cite{Ketterle} atoms depending on the atomic element and the cooling technique.

In our setup we are able to condense more than 120$\cdot$10$^{6}$ sodium atoms, which is, as far as we know, the largest condensate made with a cooling strategy based on laser cooling. In the hydrogen BEC experiment \cite{H} a cryogenic trap is used as a starting point for the evaporative cooling. This allows orders of magnitude more atoms to start with resulting in 1$\cdot$10$^{9}$ atoms in the condensate.

In our experiment, where the condensate is used as a source for superradiant scattering experiments, a high number of atoms in the BEC is essential, because the coupling of atoms to the superradiant scattered states scales exponentially with the number of atoms in the condensate. In this case a high atom number will give a strong increase in the signal to noise ratio and allows for the excitation of higher order modes. Furthermore, the high number of atoms yields the possibility to enter the hydrodynamic regime.

In this article we will describe our setup and the techniques to realize such a large condensate. We start with a description of the ultra high vacuum (UHV) system, which is essential for reaching BEC. In Sec. \ref{seclaser} the laser setup, which generates the laser light to control and probe the atoms, is described. For the creation of a BEC control of the experimental parameters is important. For this a computer-controlled system is developed, which will be discussed in Sec. \ref{secsequence}. The next section covers the Zeeman slower in the zero-crossing configuration used to slow the atomic beam down to 30 m/s. Atoms from the slowed beam are trapped in the dark-spot magneto-optical trap (MOT), as described in Sec. \ref{secMOT}. The atoms from the MOT are spin polarized in a high magnetic field before the transfer to the magnetic trap (MT). This process is discussed in Sec. \ref{secspinpol}. In Sec. \ref{secMT} we describe the MT. Sec. \ref{secRF} covers the evaporative cooling of the atoms. The large number of atoms give rise to three-body losses at the end of the evaporation. This problem is suppressed by decompressing the trap, which is discussed in Sec. \ref{secdecomp}. The imaging of the condensate is the subject of Sec. \ref{secBEC}. Finally, we will give the conclusions and the outlook of this work.

\section{Vacuum system}
\label{secvacuum}

For the realization of a large atom number BEC it is important to minimize the atomic losses during the cooling process. One of the important loss processes is the inelastic collisions with background gas atoms.  This is especially important during the evaporation, which takes a relative long time (50 s). For a lifetime of the cloud, which is in the order of the evaporation time, the background pressure must in general be below 10$^{-10}$ mbar. In our setup the background pressure is below 10$^{-11}$ mbar, which is abundantly enough, as will be shown in Sec. \ref{secMT}. This ultra-high vacuum is reached by carefully minimizing the beam load from the atomic source to the background pressure and maximizing the pumping speed, as described in the following.

\begin{figure}[th]
\centering
\includegraphics[width=0.5\textwidth]{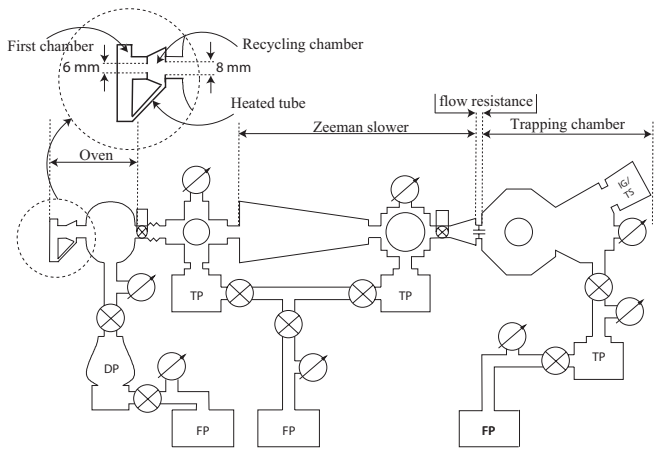}
\caption{Schematical representation of the vacuum setup (side view). From left to right the oven, a pumping section, the first part of the Zeeman slower, a pumping section, the second part of the Zeeman slower, the flow resistance, and the BEC chamber. The diffusion pump is indicated by DP, the fore-line pumps by FP, the turbo pumps by TP, and the combination of an ion getter and titanium sublimation pump by IG/TS. }
\label{vacuum}
\end{figure}

Our setup (Fig. \ref{vacuum}) consists of three main parts: the oven, the Zeeman slower and the trapping chamber. The first part is a recycling sodium oven, in which 50 g of solid sodium is heated up to 570 K.
The first chamber of the oven consists of nw-35 conflat parts. The sodium vapor leaves this chamber through a 6 mm diameter diaphragm. The sodium beam enters the recycling chamber, which consists of a conical adapter to nw-63. In this chamber, atoms outside the well-collimated beam are removed from the beam by an extra 8 mm diaphragm. To the recycling chamber, a 6 mm internal diameter tube is welded, which is connected to the bottom of the first chamber, allowing the sodium to flow back to the first chamber. The relative large internal diameter of the first part of this recirculating oven is required to limit the pressure inside this chamber, which attenuates the primary beam. The second diaphragm is heated up to 383 K due through heat conduction from other parts of the oven, which keeps the sodium in the fluid phase.
We choose not to apply any gold-coated stainless steel mesh either in this chamber or in the recirculating tube, because wetting can be problematic for sodium. At the running temperatures the 50 g sodium lasts for 5000 hours without recirculation. This is as long as we are using it until now, so no conclusion about the recirculation can be drawn at present. The oven is pumped with a 1200 l/s diffusion pump to a pressure of 4$\cdot$10$^{-8}$ mbar.

The second part of the setup is the Zeeman slower, which is used to slow down the thermal atomic beam \cite{Metcalf}. This device will be discussed in more detail in Sec. \ref{secZeeman}. Regarding the vacuum system the Zeeman slower consists of a 1900 mm long tube with a diameter of 50 mm yielding a conductivity of 8 l/s. One turbo pump is mounted between the oven and the beginning of the Zeeman slower and another pump is mounted in the Zeeman slower 1300 mm from the beginning. This configuration leads to pressure of 1$\cdot$10$^{-9}$ mbar at the point of the second turbo pump. Between the Zeeman slower and the BEC chamber a tube with a length of 120 mm and a diameter of 17 mm is mounted as a differential pumping section yielding a conductivity of 5 l/s.

\begin{figure}[th]
\centering
\includegraphics[width=0.4\textwidth]{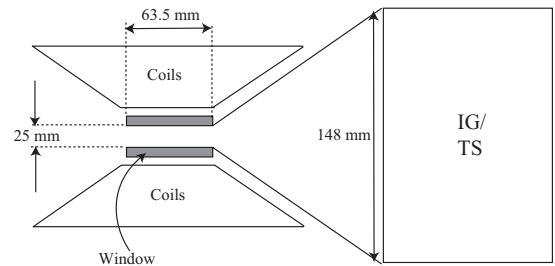}
\caption{Schematical representation of the mounting of the ion getter and titanium sublimation pump (IG/TS) (upper view).
The vacuum camber is at the point of the pump very wide to maximize the pumping speed. The diameter decreases towards the trapping region, because the coils must be as close to the center as possible to create high magnetic field gradients.}
\label{vacuum2}
\end{figure}

The BEC chamber is equipped with 9 windows for optical access. All windows are mounted with aluminum Helicoflex rings on both sides of the window to provide a leak-tight glass-to-metal sealing. The leak rate of each window is measured to be 7$\cdot$10$^{-11}$ mbar l/s. The total inflow in the BEC chamber including the beam load out of the Zeeman slower is 6$\cdot$10$^{-9}$ mbar l/s. The BEC chamber is pumped by an ion getter and titanium sublimation pump, mounted at a CF-160 port with a specified pumping speed of 1000 l/s. From the pump towards the trapping region the chamber has a conical shape to minimize the pumping resistance (Fig. \ref{vacuum2}). With this shape the actual pumping speed is likely to be close to the specified pumping speed. Therefore, the inflow of 6$\cdot$10$^{-9}$ mbar l/s is compensated at a pressure of about 6$\cdot$10$^{-12}$ mbar, which is below the limit of our pressure gauge of 1$\cdot$10$^{-11}$ mbar. The downside of this configuration is that the thickness of the window and the conical shape force us to put the coils further from the atoms leading to a reduced field gradient.

\section{Laser system}
\label{seclaser}

\begin{figure}[th]
\centering
\includegraphics[width=0.5\textwidth]{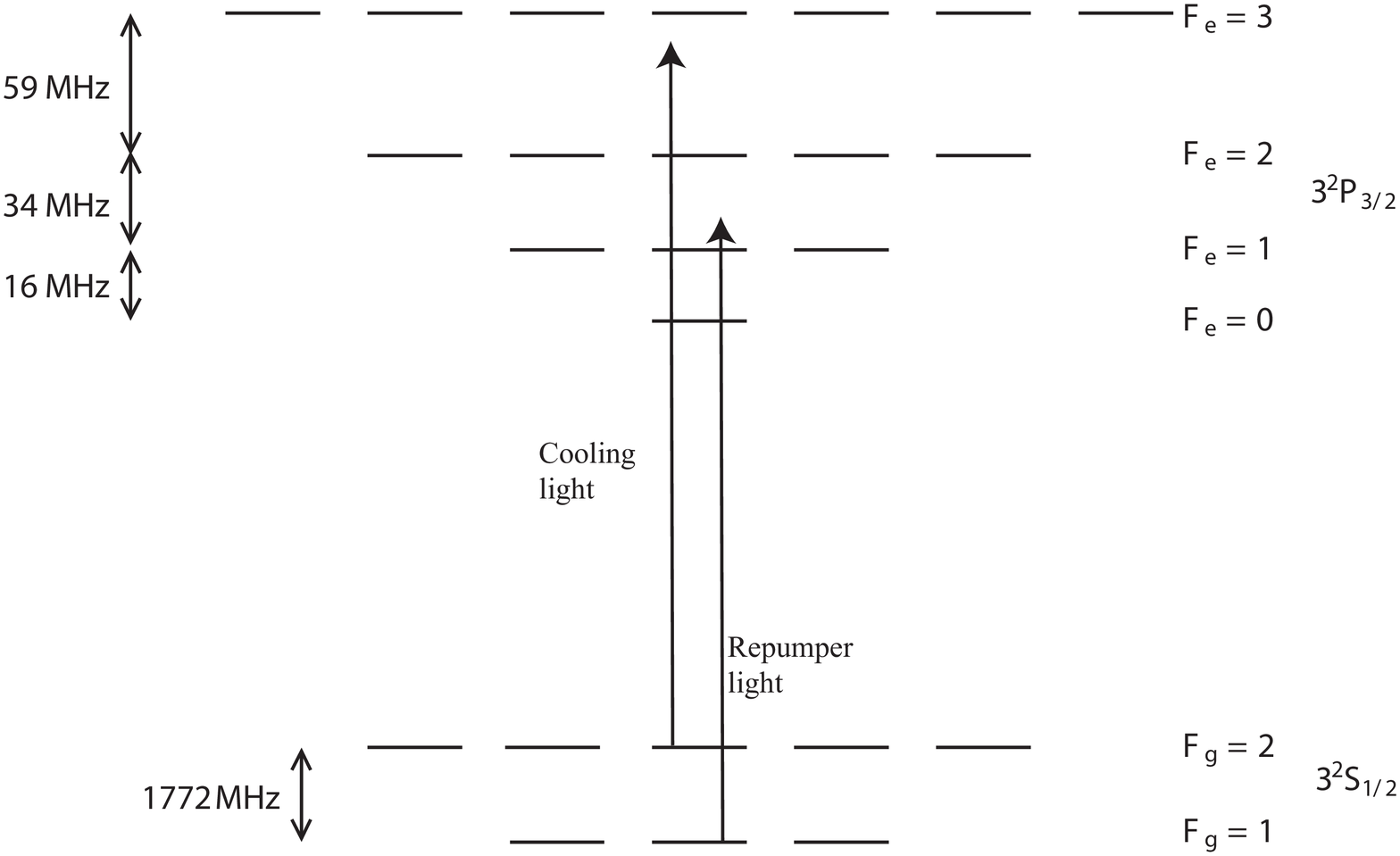}
\caption{The hyperfine structure of sodium. For laser cooling transition from the $F_{g}=2$ ground state to $F_{e}=3$ excited state is used. If an atom decays to the $F_{g}=1$ ground state, due to off resonant scattering to the $F_{e}=2$ excited state, it will be pumped back by the repumper.}
\label{levelscheme}
\end{figure}

The first step in the cooling process towards BEC is taken by laser cooling and trapping \cite{Straten}. For this we need laser light with the proper wavelength to drive a closed transition. For sodium the transition at a wavelength of 589 nm from the 3$^2$S$_{1/2}$ to the 3$^2$P$_{3/2}$ is closed (Fig. \ref{levelscheme}).  For laser cooling the hyperfine transition from the $F_{g}=2$ ground state to $F_{e}=3$ excited state is used. Due to off-resonant scattering to the $F_{e}=2$ excited state followed by decay to the $F_{g}=1$ ground state (separated by 1772 MHz) about 2 \% of the atoms will be lost from this transition. After 350 absorption-emission cycles, which is only a small fraction of the cycles needed to slow down a thermal beam, less than 0.1 \% of the atoms are still in the $F_{g}=2$ ground state. This problem is circumvented by re-exciting the atoms from the $F_{g}=1$ ground state with the so-called repump laser.

\begin{figure}[th]
\centering
\includegraphics[width=0.45\textwidth]{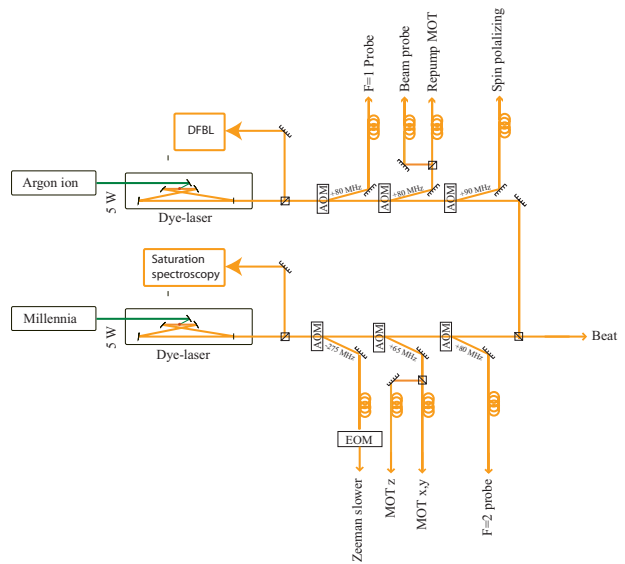}
\caption{Laser setup before entering the optical fibers. One dye-laser is used for the generation of light at the cooling transition (lower one). This light is split in four beams: one for the Zeeman slower beam, one for the MOT beams in the $x$ and $y$ direction (splitting into two separated beams is done after passing the fiber), one for the MOT beams in the $z$ direction, and one for the probe beam, used for the illumination of the atomic sample. The other dye-laser (upper one) generates light at the repump frequency. This light is also divided into four beams: one for repump light of the MOT, one for spin polarizing the sample, one for the probe beam, used for the illumination of the atomic sample, and a probe beam for the atomic beam. The frequency difference of the two lasers is monitored by beating the two lasers on a photo diode. }
\label{lasersetup}
\end{figure}

In our experiment we use for each frequency (cooling and repumping) a separated laser (Fig. \ref{lasersetup}). The advantage of this system compared to setups, where only one laser is used, is that we do not need optical power consuming elements to shift the frequency by 1772 MHz. Both laser frequencies are produced by Spectra-Physics 380D dye-lasers. The dye-laser that produces light at the laser cooling frequency is pumped with a 5 W solid state CW laser (Millennia, Spectra-Physics) and generates 700 mW light. The other dye-laser, which is used for the repumping is pumped with a 5 W argon-ion laser (serie 2060, Spectra-Physics) and generates 500 mW light. The cooling laser is locked to the cooling transition frequency by frequency modulated saturation spectroscopy \cite{demtoder}. For this a small percentage of the light is split off and used for the spectroscopy. The light is divided in two beams. From one beam the intensity is measured on a photo diode after passing a quartz sodium cell. The other beam is double passed through an acousto-optical modulator (AOM). The frequency of this AOM is modulated with 6 kHz and an amplitude of 2 MHz around a center frequency of 76 MHz. The +1 order of the frequency modulated beam passes the quartz sodium cell in opposite direction while overlapping the first laser beam. The signal from the photo diode is passed through an lock-in amplifier, where the first derivative of the signal is generated. This makes it possible to lock the laser at a zero-crossing of the signal, which corresponds to the maximum of the absorption in the sodium cell. The +1 order of the AOM is locked to the $F_{g}=2$ to $F_{e}=3$ transition. Thus the lock point of the laser is shifted by -76.0 MHz with respect to the $F_{g}=2$ to $F_{e}=3$ transition.

For locking the repump laser a Doppler-free bichromatic lock (DFBL) is used \cite{ooijen2}. This lock has the advantage that no lock-in amplifier and frequency modulation is needed.  The DFBL-lock is based on the saturation spectroscopy, in which the laser frequency is locked on the slope of a resonance peak. Locking on a slope makes the frequency sensitive for fluctuations in the laser power. In the DFBL-lock two orders of an AOM are passed through a sodium cell generating two Lamb profiles. These Lamb profiles are electronically subtracted resulting in a Doppler free signal with a possibility to lock on a zero-crossing. The lock point can be tuned by changing the AOM frequency or by using a different combination of orders of the AOM.  We use the $+1$ and $+2$ order of a 70 MHz AOM. This results in a laser frequency of -77.5 MHz with respect to the $F_{g}=1$ to $F_{e}=1$ transition \cite{ooijen2}.

The lock point of the cooling laser is in principle independent of the parameters of the laser, which makes it the frequency reference of our experiment. Although the lock of the repump laser is not sensitive for fluctuations in the total laser power, it does depend on the balance between the two orders of the AOM. This makes the lock sensitive for pointing instabilities of the laser. We monitor this shift in the repump frequency measuring the beat signal of the two lasers. Shifts in the beat frequency are manually corrected by adjusting the balance between the two orders of the AOM used for the DFBL-lock. The beat also allows us to deduce the exact frequency of the repump laser. The fluctuation of the beat frequency is less than 2 MHz, from which we conclude that the frequency fluctuations of each laser remains within 1 MHz, assuming both lasers have the same stability.

All laser beams needed for the experiment are controlled by AOM's. By adjusting the frequency and the power of the rf-input of the AOM we have full control over the frequency and the intensity of the laser beams. After passing the AOM's the light is transported through single mode polarizing preserving optical fibers (Fibercore, HB600) to the setup. The fiber has a diameter of 125 $\mu$m and a numerical aperture of 0.15.  The incouple efficiency of the fibers is between 45 and 65 \%, depending on the qualitiy of the beam shape entering the fiber. In Tab. \ref{laser} an overview of the detuning and the power after passing the fibers is given for all laser beams.

\begin{table}
\centering
\begin{tabular}{|l|l|l|l|} \hline
Purpose&Transition&Detuning (MHz) & Power (mW)\\ \hline
MOT x-y&F=2 $\rightarrow$ F=3& -11 & 80\\ \hline
MOT z&F=2 $\rightarrow$ F=3& -11 & 13\\ \hline
Zeeman slower&F=2 $\rightarrow$ F=3& -351 & 130\\ \hline
F=2 probe &F=2 $\rightarrow$ F=3& +4 & 6.0$\cdot$10$^{-4}$\\ \hline
 &  & & \\ \hline
Purpose &Transition& Detuning (MHz)& Power (mW)\\ \hline
F=1 probe&F=1 $\rightarrow$ F=1 & +2.5 & 6.0$\cdot$10$^{-4}$\\ \hline
Repumper&F=1 $\rightarrow$ F=1 & +2.5 & 1.2\\ \hline
Beam probe &F=1 $\rightarrow$ F=1& +2.5 & 1.0\\ \hline
Spin polarizing&F=1 $\rightarrow$ F=2& -3.5 & 2.0\\ \hline
\end{tabular}
\caption{Overview of the used laser frequencies and powers. The detuning in the third column is given with respect to the transition in the second column.}
\label{laser}
\end{table}

\section{Sequencing}
\label{secsequence}

For the realization of a BEC a large number of different steps have to be taken. The starting time and the duration of several of these steps is very critical. This requires a computer controlled system that can switch and adjust with $\mu$s-resolution the parameters of the experiment: the AOM's, the light and atomic beam shutters, the currents through the coils, the rf-frequency, the rf-power and the CCD-camera.

In our experiment this is done by programming a sequence on a computer and subsequently send the sequence to two National Instruments PCI-cards. The cards generate the signals that drives the setup. The digital and analog signals are produced by a PCI-6534 card and PCI-6713 card, respectively. The signals from the digital card are used for switching devices on and off at a 2$\mu$s time scale. The signals from the analog card are used for ramping the magnetic trap up and down. The rf-field for the evaporative cooling is generated by a direct digital synthesizer (DDS).

In Tab. \ref{sequence} an overview of the timing of the different steps during the cooling towards BEC and the imaging of the BEC are summarized. A more thorough discussion of each step will follow in the remaining part of this article.

The clock of the experiment is an frequency shifting key (FSK) clock. The frequency of this clock is controlled by one of the digital lines of the experiment. The sequence can thus be slowed down by a factor of 100 during times where tight timing is not required, conserving time and memory and allowing us to reach 2$\mu$s resolution in experiments running for more than 200 s.

\begin{table*}
\centering
\begin{tabular}{|l|l|l|l|} \hline
Process&Duration&Number of particles&Temperature ($\mu$K)\\ \hline
Loading dark-spot MOT& 6 s & 2.0$\cdot$10$^{10}$ & 320\\ \hline
Turn off MOT magnetic field & 1 ms &  2.0$\cdot$10$^{10}$ & 320\\
and the repump laser beam& & & \\ \hline
Turn on the magnetic field for the & 500 $\mu$s &  2.0$\cdot$10$^{10}$ & 320\\
spin polarizing &  & & \\ \hline
Apply spin polarizing laser pulse & 500 $\mu$s &  2.0$\cdot$10$^{10}$ & 320\\ \hline
Apply depump laser pulse &500 $\mu$s &  2.0$\cdot$10$^{10}$ & 320\\ \hline
Turn on MT/Turn off laser beams &2 ms &  & \\ \hline
Stabilization of the MT &4 s &  1.4$\cdot$10$^{10}$ & 350\\ \hline
Evaporative cooling in a compressed trap& 42 s& & \\ \hline
Decompression and evaporative cooling & 2 sec & & \\ \hline
Evaporative cooling in a decompressed trap & 6 sec & & \\ \hline
Turn off the MT&750 $\mu$s&  120$\cdot$10$^{6}$ & 0.15\\ \hline
Time of flight & 0 to 250 ms & 120$\cdot$10$^{6}$ & 0.15\\ \hline
Imaging of the atoms &400 $\mu$s&  120$\cdot$10$^{6}$ & 0.15\\ \hline
Escape of the atoms &1 s &  & \\ \hline
Imaging of the image laser beam &400 $\mu$s& &\\ \hline
Wait time & 1 s & & \\ \hline
Background image &400 $\mu$s& &\\ \hline
\end{tabular}
\caption{Overview of the BEC sequence. The duration, the number of particles and the temperature of the cloud are given at different instants in the BEC process. During the evaporation the number of particles and the temperature is changing, therefore the values for these two parameters during this process are not given.}
\label{sequence}
\end{table*}

\section{Zeeman slower in zero crossing configuration}
\label{secZeeman}

Our recycling oven produces a sodium flux of 3$\cdot$10$^{12}$ atoms/s at a temperature of 570 K. This flux is determined by measuring the absorption from a beam probe at the end of the Zeeman slower. If we assume that the velocity distribution of the atoms out of the oven obeys the Maxwell-Boltzmann distribution at a temperature of 570 K, only 0.02 \% of the atoms have a velocity smaller than 30 m/s, which is the capture velocity of our trap. The loading of the trap can be increased significantly by predecelerating the beam in a Zeeman slower. In a Zeeman slower the atoms are decelerated by a counter propagating laser beam, while the changing Doppler shift is compensated by a changing Zeeman shift due to an inhomogeneous magnetic field along the deceleration path. In the Zeeman slower the longitudinal velocity decreases, but there is no cooling in the radial direction. The loss of atoms due to this increasing divergence can be compensated by placing the end of the Zeeman slower as close to the trapping chamber as possible. In this situation it is complicated to install a vacuum pump between the Zeeman slower and the trapping chamber. Therefore, our Zeeman slower is divided in two parts, where in between the two parts a pump is placed. The pumping section is passed by the atoms with a velocity of 200 m/s, so that the beam divergence is still small at this point.

Other ways of minimizing the losses due to the divergence is by using 2D optical molasses \cite{Straten}, a magneto-optical lens \cite{lens} or a magneto-optical compressor \cite{compressor} at the end of the Zeeman slower. All of these options have proved to be efficient; however, they need extra laser beams, which makes them in day-to-day operation less favorable compared to our solution.

\begin{figure}[th]
\centering
\includegraphics[width=0.45\textwidth]{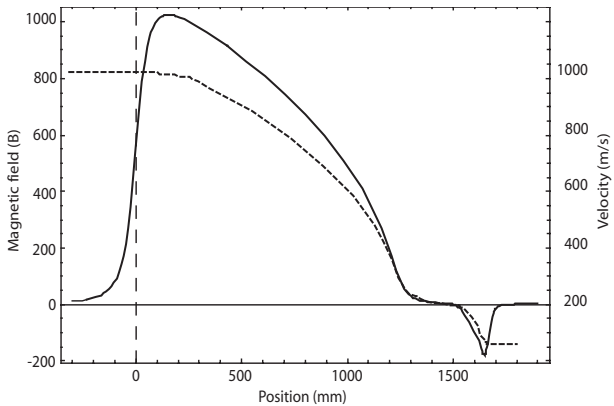}
    \caption{The solid line shows the calculated magnetic field as a function of the distance from the beginning of the         Zeeman slower. Between the 1200 and 1500 mm a pumping section is implemented resulting in a zero magnetic field. The dotted line shows the velocity of the atoms during the slowing. The trapping region is at 1900 mm from the beginning of the Zeeman slower.}
    \label{slower}
\end{figure}

In Fig. \ref{slower} the calculated magnetic field in the Zeeman slower as a function of the distance from the beginning of the Zeeman slower is given (solid line). The field starts at 1000 G for a capture velocity of 950 m/s. In the Zeeman slower the atoms pass a region with a zero magnetic field. In this region the energy splitting of the magnetic sublevels is not large enough to prevent off-resonant scattering to the $F_{e}=2$ excited state followed by decay to the $F_{g}=1$ ground state.  Therefore, a repumper is needed to prevent losses from the cooling process during the passage of the pumping section. This repump frequency is generated by passing the laser beam through an electro-optical modulator (EOM) operating at 1720 MHz. The side band generated with the EOM contain 10 \% of the total intensity. The EOM is placed after the optical fiber because the EOM crystal disturbs the beam profile to much to get sufficient transmission through the fiber ($>$45\%).

Figure \ref{slower} also shows the calculated velocity of the atoms passing the Zeeman slower (dotted line). We can see that the final velocity (30 m/s) is not reached until close to the trapping region, which is a great benefit for the number of particles in the slowed beam. The slow atom flux is measured with the beam probe to be 7$\cdot$10$^{9}$ atoms/s.
Taken the divergence of the atomic beam due to the slowing into account the slowing efficiency is determined to be around 10 \%. We have not yet fully tracked down the reason for this sub-optimal performance.

\section{The dark-spot magneto optical trap}
\label{secMOT}

With atoms slowed down to 30 m/s, a dark-spot MOT is loaded for 6 s. In the dark-spot MOT configuration the repump beam has a black spot in the center of the beam. The spot is imaged at the MOT, where it has a diameter of 12 mm. The atoms in the center of the dark-spot MOT are optically pumped to the $F_{g}=1$ ground state due to the presence of only cooling light \cite{darkmot}. This prevents the density limitation due to intra-MOT collisions in a bright MOT \cite{intramot1,intramot2}. The light of the dark-spot MOT consists of three retro-reflected laser beams (MOT beams) with beam diameter of 25 mm and a detuning of -11 MHz with respect to the $F_{g}=2$ to $F_{e}=3$ transition, and one beam (repump beam) with a detuning of +2.5 MHz with respect to the $F_{g}=1$ to $F_{e}=1$ transition (see Tab. \ref{laser}).

The quadrupole magnetic field needed for the MOT is generated by two anti-Helmholtz coils producing a field gradient of 5 G/cm at a current of 15 A. A typical dark-spot MOT in our experiments consists of 2.0$\cdot$$10^{10}$ atoms with a temperature of $320$ $\mu$K at a peak density of about 1.0$\cdot$10$^{11}$ atoms/cm$^{3}$\cite{ooijen}.
Note, that due to the slow atom flux of 7$\cdot$10$^{9}$ atoms/s the MOT loads in a few seconds.

\section{Spin polarizing in a high magnetic field}
\label{secspinpol}

After loading the dark-spot MOT the atoms are transferred to the MT for the evaporative cooling. The efficiency of this transfer depends on the distribution of the atoms over the magnetic sublevels of the trapped state. The sodium atoms trapped in a dark-spot MOT occupy the $F_g=1$ ground state. The homogeneous distribution over the three magnetic sublevels, makes that only 1/3 of the atoms are in the "low field seeking" state ($F_{g}=1, M_{g}=-1$), which can be trapped in the minimum of the magnetic trap. The $F_{g}=2, M_{g}=2$ state, which can also be trapped in the minimum of the magnetic trap with two times stronger confinement is not investigated. It is harder to reach BEC with atoms in this state, since only a pure $F_{g}=2, M_{g}=2$ state is stable against spin changing collisions. Therefore, all the atoms in other $F_{g}=2$ ground states must be removed from the trap before evaporation \cite{Na2}.

\begin{figure}[th]
    \centering
        \includegraphics[width=0.5\textwidth]{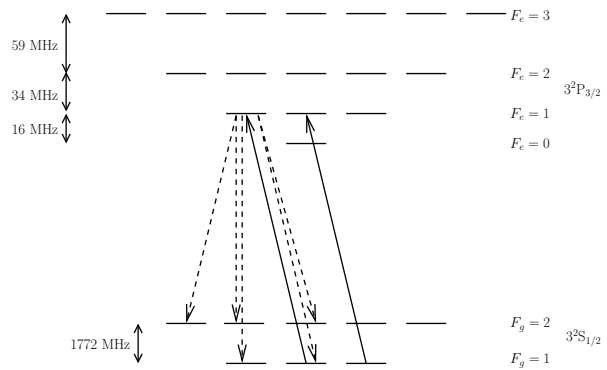}
    \caption{The principle of spin polarizing. The atoms in the $F_{g}=1$ ground state are optically pumped to the excited state by a laser beam with $\sigma_-$ polarization (solid arrows). The atoms in the $F_{e}=1$ excited state can decay back to the $F_{g}=1, M_{g}=-1$ ground states, making them spin polarized. However, also decay to the other $F_{g}=1$ ground states is possible, which makes another absorption-emission cycle needed. An atom in the $F_{e}=1$ excited state can also decay to the $F_{g}=2$ ground state. From this state the atoms are depumped by the MOT laser beams back to the $F_{g}=1$ ground state. The different decay possibilities from one excited state are shown indicated with dashed arrows. The depump transition are not shown.}
    \label{spinpolprincipe}
\end{figure}

The transfer efficiency from the MOT to the MT of 1/3 can be increased by spin polarizing the sample before loading it into the MT. This can be done by applying a short laser pulse with the proper polarization in combination with a small magnetic field to define the spin state with respect to the magnetic trap (Fig. \ref{spinpolprincipe}). Spin polarization is a commonly used technique in  BEC experiments, for example in the experiments with Rb \cite{Rb}, He \cite{He}, and Ne \cite{Ne}. In these experiments transfer efficiencies up to 0.70 are reached. To our knowledge in sodium BEC experiments the efficiency has never exceeded 0.35 \cite{Na2} due to the large number of transitions lying very close to each other in frequency in combination with the high optical density. The final degree of polarization is determined by the balance between the optical pumping to the polarized state and the depolarization. The depolarization is caused by off-resonant scattering of the spin-polarization beam by atoms, which are already polarized, and the reabsoption of spontaneous emitted photons. In sodium MOTs used in BEC experiments the optical density is so high, that these are strong depolarization mechanisms. For sodium it is not possible to circumvent this problem using a larger detuning due to the large number of optical transitions, which are close to each other in frequency.

\begin{figure}[th]
\centering
\includegraphics[width=0.42\textwidth]{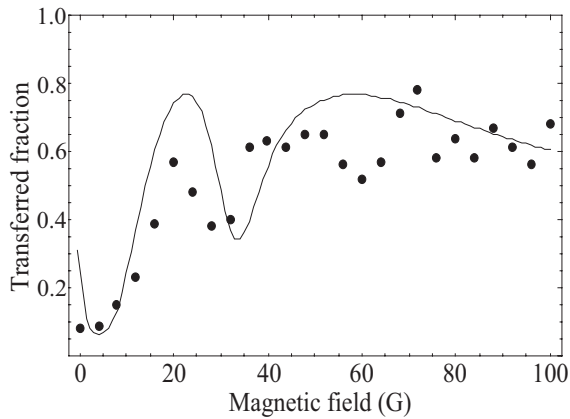}
    \caption{Transfered fraction of atoms from the MOT to the MT as a function of the applied magnetic field.}
    \label{magneetveld}
\end{figure}

In our experiment we spin polarize the atoms in a high magnetic field. The high magnetic field induces large level splittings, which makes it easier to drive only the transition needed for the polarization process. Furthermore, due to the large level splittings the optical density of the sample decreases significant, which makes the re-absorption of spontaneous emitted photons no longer problematic \cite{spinpol}. The magnetic field is created by a combination of the fields generated by the pinch and bias coils of the magnetic trap. The details of this trap will be discussed in Sec. \ref{secMT}. With these coils we can tune the field from 0 to 125 G with a accuracy of 1 G. Directly after the MOT magnetic field and the repump laser beam are turned off the magnetic field of the spin polarizing is turned on. Note, that the MOT laser beams are still on. It takes about 500 $\mu$s for the field to stabilize. After this time the spin polarizing beam is turned on for 500 $\mu$s. During spin polarizing the MOT beams pump the atoms decaying to the $F_{g}=2$ ground state back to the $F_{g}=1$ ground state. After spin polarizing these beams are kept on for another 500 $\mu$s to make sure all the atoms end in the $F_{g}=1$ ground state.  In Fig. \ref{magneetveld} the transfer efficiency from the MOT to the MT is given as a function of the applied magnetic field. This figure clearly shows that spin polarizing in a high magnetic field gives a high and stable transfer efficiency around 0.7, which is comparable to other spin polarization experiments. The solid line in Fig. \ref{magneetveld} gives the calculated transfer efficiency. In this calculation the optical Bloch equations are solved for the multilevel structure of sodium in a magnetic field. After elimination of the population of the excited state rate equations for transitions from one ground state to another ground state are obtained. These rate equation form the bases for the calculation. A more detailed description of the model is given in Ref. \cite{spinpol}.

\section{The magnetic trap}
\label{secMT}

For trapping the atoms during the evaporative cooling we use a magnetic trap in the cloverleaf configuration \cite{cloverleaf}. The trap is schematically shown in Fig. \ref{mt}. The two pinch coils generate a curvature in the magnetic field in the $z$-direction, which causes confinement along this direction. The offset in the magnetic field is compensated by the bias coils producing a field in the opposite direction. The gradient coils produces a quadrupole field for the radial confinement. The advantage of the cloverleaf trap is the full 360$^{\circ}$ optical access.

\begin{figure}[th]
\centering
\includegraphics[width=0.45\textwidth]{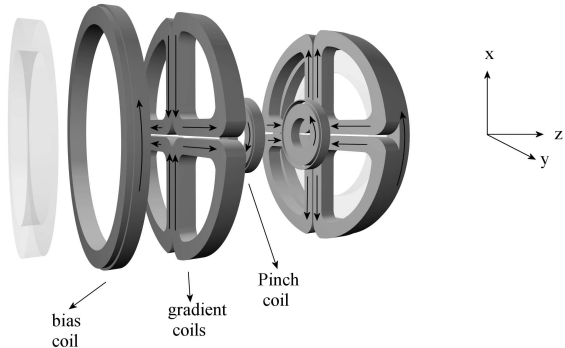}
    \caption{Schematic view of the cloverleaf trap. From left to right we see the MOT coils (transparent), the bias coils, the set of four gradient coils and the pinch coils. Furthermore we can see the fine tune coils around the pinch coils.}
    \label{mt}
\end{figure}
We operate the gradient and the pinch/bias coils at 300 A and 200 A, respectively. The current of 300 A through the gradient coils yields a magnetic field gradient of $118$ G/cm in the radial direction. The curvature field in the axial direction is $42$ G/cm$^2$ at a current of 200 A through the pinch/bias combination. The minimum of the field is 3.4 G. This leads to trap frequencies in the axial and radial direction of $\nu_{z}=16$ Hz and $\nu_{\rho}=99$ Hz. The stability of the minimum of the magnetic field is essential for a reproducible evaporation. The minimum is generated by subtracting the two relative large magnetic fields of the pinch and the bias coils, which makes it very sensitive for the current through this coils as well as the position of the coils.

To minimize the first effect the current through the pinch and bias coils is produced by the same power supply. In parallel with the bias coils a shunt is placed, to fine tune the current through the bias coils. The second effect is minimized by gluing all the windings of the separated coils together with Loctite, subsequently both sets of coils are fixed in polyester boxes. The coils are made of hollow, rectangular copper wires, through which cooling water is pumped. The temperature of the cooling water is stabilized within 1 $^0$C. With these precautions the stability of the field minimum of our trap is 0.4 mG. This is determined by measuring the stability in the number of particles in the BEC, when the atoms are cooled very close to the bottom of the trap. The stability of the trap is 500 Hz, which corresponds to 0.4 mG.

The switch-on time of the coils is less than 2 ms. Most information from the cold atoms is obtained from time of flight images of the expanding cloud. Therefore, the expansion of the cloud should be dominated by the dynamics of the cloud and not by the remaining magnetic field. This makes it important that the magnetic field is switched off fast on the time scale of the trap period (less than 1 ms). To meet these requirements all the sets of coils are equipped with a fast switch. Our switch consist of a resistor and a diode, which are placed in parallel with the coils, and an insulated gate bipolar transistor (IGBT, Dynex DIM600BSS17-E000) in serial to the coils. To switch off the magnetic field the IGBT is opened. This is done relatively slowly (50 $\mu$s) on the time scale of the IGBT to limit the peak voltage arising during the turn-off to 1200 V. This limitation does not increase the switch-off time, because this is determined by the self induction of the coils. The energy that is released from the coils is absorbed in the resistor. Because a diode is placed in serial with the resistor, no current flows through the resistor during normal operation of the magnetic trap. To suppress the limitations in the switch-off time due to eddy currents the amount of metal between the coils is minimized. This is achieved by mounting large windows (radius of 63.5 mm) in the vacuum chamber in the $z$-direction (Fig. \ref{vacuum2}). The radius of the windows is twice as big as the outer radius of the pinch coil. The switch-off time of the coils is less than 300 $\mu$s (1/e time). In the experiment we have increased the time to 750 $\mu$s, because a switch-off time of 300 $\mu$s generates repulsive forces between the coils large enough to crack the polyester box in which they are mounted.

After spin polarizing, as described in the previous section, the gradient coils are turned on and the shunt is closed, which changes the high bias field used for the spin polarizing into a curvature field. The strength of the trapping potential of the MOT and the MT must be matched, so that heating during the transfer is avoided. When the MT is much stronger than the MOT, this can be done by turning the MT on slowly. However, in our experiment we can turn on the magnetic trap immediately at the full strength without problems of mode matching the distributions of the atoms in the MOT with the distribution of the atoms in the MT. In the MT we can load 1.4$\cdot$10$^{10}$ atoms at a temperature of 350 $\mu$K with a peak density of 5.0$\cdot$10$^{10}$ atoms/cm$^{3}$.

\begin{figure}[th]
\centering
\includegraphics[width=0.45\textwidth]{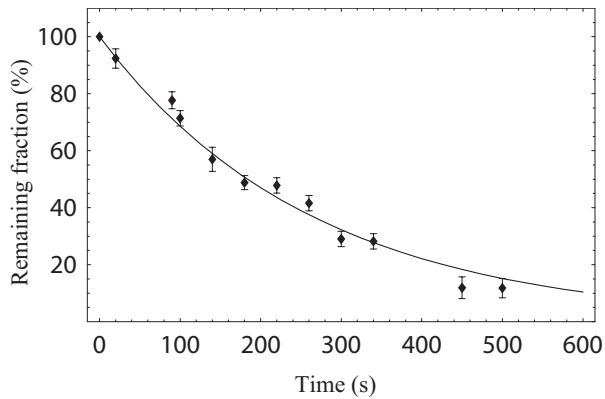}
\caption{Number of atoms in the MT as a function of the storage time. The solid line is fit of the data to an exponential decay yielding a lifetime of 260 s.}
\label{lifetime}
\end{figure}

In Fig. \ref{lifetime} the amount of atoms in the MT is plotted as a function of the storage time in the MT. From the fit of the data to an exponential decay the lifetime of the atoms in the MT is determined to be 260 s. This is long enough for the evaporative cooling, which takes only 50 s. Therefore, we loose only $\approx$ 20\% of the particles due to the finite pressure.

\section{Evaporative cooling}
\label{secRF}

The atoms in the magnetic trap are cooled to degeneracy by evaporative cooling. The basic principle of this technique is that the atoms with an energy higher than the average kinetic energy are removed from the sample by spin-flipping them to an untrapped state, followed by rethermalization of the remaining atoms due to elastic collisions \cite{evap1,evap2}. The energy difference between the trapped and the untrapped state is given by $\mu_BB/2$  corresponding to a frequency of 50 MHz at 70 G. The field for spin-flipping is generated by a single-turn rf coil (diameter of 21 mm) driven by a 10 W amplifier. The coil is placed outside the vacuum system between the window and the pinch coil at 40 mm distance from the atom cloud. The rf-frequency is ramped down from 50 MHz to 2.5 MHz in 50 s. The frequency in the first 42.5 s is linearly ramped down with a rate of 1.08 MHz/s; in the last 7.5 s the ramp rate is decreased to 200 kHz/s. The rf-power is kept at a constant level of 10 W during the ramp. The rf-field generates an oscillating magnetic field with an amplitude in the order of 10 mG. The probability to make a transition from the trapped to untrapped state equals $\approx$0.5 at the beginning of the evaporation and will increase rapidly with decreasing temperature \cite{ooijen}.

\begin{figure}[th]
\centering
\includegraphics[width=0.45\textwidth]{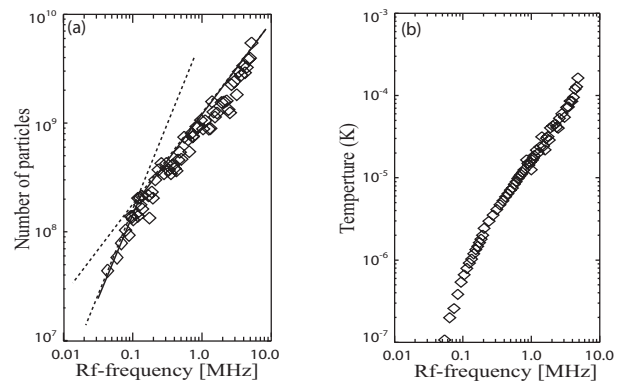}
    \caption{(a) Number of atoms as a function of the rf-frequency with respect to the bottom of the trap. At frequencies below 150 kHz we see that the efficiency parameter $\alpha$ decreases, as indicated with the two dashed lines. (b) The temperature of the cloud as a function of the rf-frequency. BEC is reached at a temperature of 150 nK.}
    \label{rf}
\end{figure}

For an evaporation time of 50 s we can condense 35$\cdot$10$^{6}$ atoms in the setup. Even in 15 s we can reach condensation but with significant less atoms. In Fig. \ref{rf}a the number of atoms is plotted as a function of the rf-frequency relative to the bottom of the trap for a 50 s ramp. During the evaporation the density increases to 3$\cdot$10$^{14}$ atoms/cm$^{3}$. With a 3-body collision rate of $G_3$ = 1.1$\cdot$10$^{-30} $cm$^{6}$/s \cite{G3} this yields a large loss rate of $G_3 \cdot n^2$ = 0.1 /s. The loss rate due to two-body collisions is given by $G_2$$\cdot$$n$, with $G_{2}$ = 6$\cdot$10$^{-17}$ cm$^{3}$/s \cite{G2} and $n$ is the density of the sample. This loss rate is small compared to the 3-body lose rate. For the evaporation an efficiency parameter $\alpha$ can be defined as $\alpha=\frac{\dot{T}/T}{\dot{N}/N}$ \cite{Ketterleevap}. The high loss rate due to 3-body collisions results in a decreasing $\alpha$, as shown in the last part of the evaporation ramp (below 150 kHz). In Fig. \ref{rf}b the temperature of the cloud is plotted as a function of the rf-frequency with respect to the bottom of the trap. BEC is reached at a temperature of 150 nK and a rf-frequency of 50 kHz.

\section{Evaporative cooling in a decompressed magnetic trap}
\label{secdecomp}

\begin{figure}[th]
\centering
\includegraphics[width=0.45\textwidth]{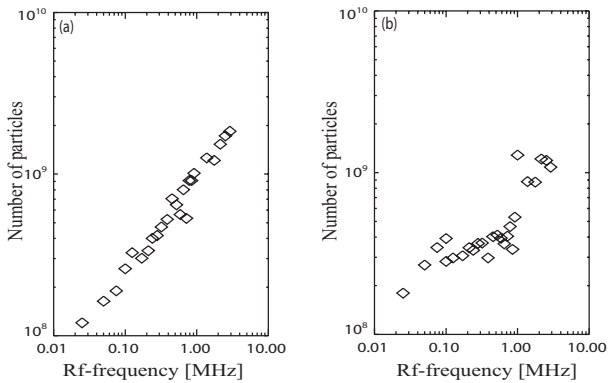}
\caption{The number of particles during the evaporation in a decompressed trap. The left picture is for the evaporation in the radial decompressed trap, resulting in 60$\cdot$10$^{6}$ atoms in the condensate. The right picture is for the axial decompressed situation leading to a BEC with 120$\cdot$10$^{6}$ atoms.}
\label{decomp}
\end{figure}

To reduce the losses due to 3-body collisions the trap is decompressed after 42 seconds of evaporation, corresponding with a rf-frequency of 2.04 MHz with respect to the bottom of the trap. At higher frequencies and thus higher temperatures 3-body losses are not the limiting loss process, since the density is always below 2$\cdot$10$^{13}$  atoms/cm$^{3}$ yielding a loss rate of 4$\cdot$10$^{-4}$ /s. For the decompression the trap frequencies can be lowered in either axial or radial direction. In Fig. \ref{decomp} the results of those measurements are shown, where the radial and the axial decompression reduces the trap frequencies from 99 to 50 Hz and from 16 to 4 Hz, respectively. Note, that the radial decompression takes place in 2 directions instead of only 1 direction in the axial decompressed case. This means that the decrease in density is the same in the two cases. The losses compared to the compressed trap are strongly reduced by decompression. The evaporation in the radial decompressed trap results in 60$\cdot$10$^{6}$ atoms in the BEC. This is increase of almost a factor of two compared to the compressed trap. The number of atoms can even be increased by a factor of four by evaporation in an axial decompressed trap. In this elongated trap we can condense 120$\cdot$10$^{6}$ atoms. This is as far as we know the largest atom number in a BEC starting from an optical trap.

The fact that the axial decompression works better than radial decompression is a strong indication that the collisional density is high enough to generate avalanches, which heats the atoms in the trap strongly \cite{avalanches}. In a collisional opaque sample evaporated atoms have a larger probability to escape from an elongated cigar shaped trap without creating an avalanche than from a more spherical symmetric trap. The collisional opacity is given by $\left\langle nr \right\rangle \sigma_s$, with $\sigma_s=8\pi a^2$ the s-wave cross section, $a$ the scattering length, and $\left\langle nr \right\rangle$ the average column density. In our experiment the collisional opacity is 1.7, which is significant higher than the critical value of 0.693 \cite{avalanches} .

\section{Probing of the Bose Einstein Condensate}
\label{secBEC}

The atoms are probed by absorption imaging. For this purpose the atoms are overlapped with a collimated laser beam aligned perpendicular to the long axis of the trap. The shadow caused by the cloud is imaged on the CCD camera (Apogee AP1E, Kodak KAF-0401E chip). To obtain information about the momentum distribution of the atoms, the atoms are probed after a time of flight after switching off the trap. In our experiment the BEC is typically imaged after 80 ms time of flight, when the optical density is between 3 and 5. The probe beam is 2.5 MHz detuned from the $F_{g}=1$ ground state to the $F_{e}=1$ excited state transition (Tab. \ref{laser}). The atoms can also be probed at the $F_{g}=2$ ground state after optical pumping to this state, but this did not improve the quality of the pictures. The length of our probe beam is 400 $\mu$s, which is long enough for a good contrast and short enough to avoid heating of the cloud.

To extract the absorption from the image we take after the first image at 1 sec intervals two additional images: one of the probe beam without atoms and one of the background. The absorption is determined by taking the ratio of the image with atoms and the image without atoms, after first subtracting the background image from both.

The absorption profile is fitted with a density profile, which is the sum of Thomas-Fermi profile for the condensed atoms and a Gaussian profile for the thermal atoms. From the fit three parameters for the condensate are extracted: the absorption at the center of the profile due to the condensate ($A$), and the width of the Thomas-Fermi profile in the $r$- and $z$-direction ($\sigma_r$ and $\sigma_z$). The number of condensed atoms can be determined in two ways: either by the total absorption or by the width of the Thomas-Fermi profile \cite{verenna}. In the first case the number of condensed atoms becomes $N=(8\pi/15)n_0\sigma_r^2\sigma_z$, where the center density is given by $n_0=(3A/4\sigma r)$, with $r$ the radius of the BEC after expansion and $\sigma$ the absorption cross section. For the $F_{g}=1$ ground state to the $F_{e}=1$ excited state transition the absorption cross section is given by $\sigma$ = $(3\lambda^2/2\pi) C_{eq}$ and the square of the Clebsch-Gordan coefficient becomes $C_{eq}=5/18$. The second method of extracting the number of atoms from the images is to determine the chemical potential from the width of the profile. This is given by the relation:
\begin{equation}
\mu=\frac{1}{2}\frac{m\omega_r^2}{1+\omega_r^2\sigma_r^2 t^2},
\end{equation}
with $\mu$ the chemical potential, $\omega_r$ the radial trap frequency and $\sigma_r$ the Thomas-Fermi radius after the expansion. The number of particles in the BEC can then be determined with the following equation:
\begin{equation}
\mu^{5/2}=\frac{15\hbar^2 m^{1/2}}{2^{5/2}}N\bar{\omega}^3a,
\end{equation}
with $a$ scattering length, $N$ the number of atoms in the BEC,  and $\bar{\omega}$ the average trap frequency \cite{verenna}. When the used time of flight is more than 30 ms both methods give the same results within the measurement uncertainty. After expansions shorter than 30 ms the high optical density makes the first method less reliable. In our experiment we use the chemical potential for determining the number of particles (second method), because this is insensitive for fluctuations in the probe frequency, while optical density is.

\begin{figure}[th]
\centering
\includegraphics[width=0.45\textwidth,bb=0 0 433 334]{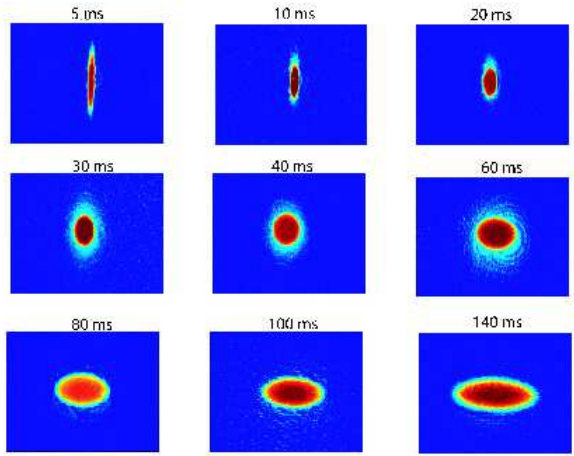}
    \caption{Absorption images as a function of the time of flight. Above each picture the time of flight is given. It is clearly shown that the aspect ratio changes from the shape dominated by the trap potential (elongated in the vertical direction) to the shape dominated by the mean field energy (elongated in the horizontal direction). }
    \label{TOF}
\end{figure}

With the decompression, as described in the previous section, we can condense 120$\cdot$10$^{6}$ atoms at a temperature of 150 nK and a density of 2.8$\cdot$10$^{14}$ atoms/cm$^{3}$. Axial and radial size of the condensate is approximately 1 mm and 20 $\mu$m, respectively. Evidence for the phase transition from a thermal gas to a BEC is the changing aspect ratio as a function of the time of flight (Fig. \ref{TOF}). After a short time the cloud of atoms still has the shape of the atom distribution in the magnetic trap, which is elongated in the vertical direction in the pictures. This aspect ratio changes due to the mean field energy, which causes the cloud to expand more rapidly in the direction of the strongest confinement. This effect can been seen clearly in our measurements.

\section{Conclusion and outlook}

To summarize, we have realized a BEC of sodium containing 120$\cdot$10$^{6}$ atoms. This is as far as we know the largest condensate starting from an optical trap. The large number of particles is mainly due to two reasons compared to similar experiments. First, we start with a large number of atoms in the MT due to spin polarizing the sample before we load the atoms in the MT. Secondly, we loose less than a factor of 120 in the number atoms during the evaporation. This relatively small loss is caused by suppression of the 1,2,3-body losses during the evaporation. The 1-body losses are strongly reduced due to the low background pressure, resulting in a lifetime of the atoms in the MT of more than a factor 5 longer than the evaporation time. Furthermore, the 2-body losses and especially the 3-body losses are suppressed by lowering the density of the cloud due to the decompression during the evaporation.

The large number of atoms provides us the possibility to do experiments such as superradiant scattering with a better signal to noise ratio. Furthermore, the large number of atoms in combination with the low trap frequencies gives rise to a high volume condensate. This in combination with the low density in the condensate compared to similar experiments yields us the possibility to enter the hydrodynamic regime.

\section{Acknowledgments}

The authors thank J. van de Weg and F. Ditewig for their contributions during the construction of the experimental setup.


\begin{thebibliography}{99}

\bibitem{Rb} M.H. Anderson, J.R. Ensher, M.R. Matthews, C.E.
Wieman, and E.A. Cornell, Science {\bf 269}, 198 (1995).

\bibitem{Na} K.B. Davis, M.-O. Mewes, M.R. Andrews, N.J. van
Druten, D.S. Durfee, D.M. Kurn, and W. Ketterle, Phys. Rev. Lett.
{\bf 75}, 3969 (1995).

\bibitem{Li} C.C. Bradley, C.A. Sackett, J.J. Tollett, and R.G.
Hulet, Phys. Rev. Lett. {\bf 75}, 1687 (1995); C.C. Bradley, C.A.
Sackett, and R.G. Hulet, Phys. Rev. Lett. {\bf 78}, 985 (1997).

\bibitem{Hansch} W. H\"{a}nsel, P. Hommelhoff, T.W. H\"{a}nsch, and J. Reichel, Nature {\bf 413}, 498 (2001).

\bibitem{Ketterle} M.W. Zwierlein, C. A. Stan, C.H. Schunck, S.M.F. Raupach, A.J. Kerman, and W. Ketterle,
 Phys. Rev. Lett. {\bf 92}, 120403 (2004).

\bibitem{H} D.G. Fried, T.C. Killian, L. Willmann, D. Landhuis, S.C. Moss, D. Kleppner,
T.J. Greytak, Phys. Rev. Lett. {\bf 81}, 3811 (1998).

\bibitem{Metcalf} W.D. Phillips, and H.J. Metcalf, Phys. Rev. Lett. {\bf 48}, 596 (1982).

\bibitem{Straten} H.J. Metcalf, and P. van der Straten, \textit{Laser Cooling and Trapping} (Springer 1999).

\bibitem{demtoder} W. Demt\"{o}der, Laser Spectroscopy, \textit{Basics Concepts and Instrumentation} (Springer, Berlin 1996).

\bibitem{ooijen2} E.D. van Ooijen, G. Katgert, and P. van der Straten, Appl. Phys. B {\bf79}, 57 (2004).

\bibitem{lens} A. Koolen, \textit{Dissipative Atom Optics with Cold Metastable Helium Atoms} Ph.D. thesis (Eindhoven University 2000).

\bibitem{compressor} E. Riss, D.S. Weiss, K.A. Moler and S. Chu,
Phys. Rev. Lett. {\bf 64}, 1658 (1990).

\bibitem{darkmot} W. Ketterle, K.B. Davis, M.A. Joffe, A. Martin, and D.E. Pritchard,
Phys. Rev. Lett. {\bf70}, 2253 (1993).

\bibitem{intramot1} M. Prentiss, A. Cable, J.E. Bjorkholm, S. Chu, E.L. Raab, and
D.E. Pritchard, Opt. Lett. {\bf13}, 452 (1988).

\bibitem{intramot2} T. Walker, D. Sesko, and C. Wieman,
Phys. Rev. Lett. {\bf64}, 408 (1990).

\bibitem{ooijen} E.D. van Ooijen, \textit{Realization and Illumination of Bose-condensed Sodium Atoms}
Ph. D. thesis (Utrecht University 2005).

\bibitem{Na2} Z. Hadzibabic, S. Gupta, C.A. Stan, C.H. Schunck, M.W. Zwierlein, K. Dieckmann, and
W. Ketterle, Phys. Rev. Lett {\bf 91}, 160401 (2003).

\bibitem{He} F. Pereira Dos Santos, J. Leonard, J. Wang, C.J. Barrelet,
F. Perales, E. Rasel, C.S. Unnikrishnan, M. Leduc, and C. Cohen-Tannoudji,
Phys. Rev. Lett. {\bf 86}, 3459 (2001).

\bibitem{Ne} S.J.M. Kuppens, J.G.C. Tempelaars, V.P. Mogendorff, B.J. Claessens,
H.C.W. Beijerinck, and E.J.D. Vredenbregt, Phys. Rev. A {\bf 65}, 023410 (2002).

\bibitem{spinpol} K.M.R. van der Stam, A. Kuijk, R. Meppelink, J.M. Vogels, and P. van der Straten,
Phys. Rev. A {\bf 73}, 063412 (2006).

\bibitem{cloverleaf} M.-O. Mewes, M. R. Andrews, N. J. van Druten, D. M. Kurn, D. S. Durfee, and W. Ketterle,
Phys. Rev. Lett. {\bf 77}, 416 (1996).

\bibitem{evap1} H.F. Hess, Phys. Rev. B {\bf 34}, 3476 (1986).

\bibitem{evap2} O.J. Luiten, M.W. Reynolds, and J.T.M. Walraven, Phys. Rev. A {\bf 53}, 381 (1996).

\bibitem{G3} D.M. Stamper-Kurn, M.R. Andrews, A.P. Chikkatur, S. Inouye, H.-J. Miesner, J. Stenger, and W. Ketterle,
Phys. Rev. Lett {\bf 80}, 2027 (1998).

\bibitem{G2} H.M.J.M. Boesten, A.J. Moerdijk, and B.J. Verhaar,
Phys. Rev. A {\bf 54}, 29 (1996).

\bibitem{Ketterleevap} W. Ketterle, and N.J. van Druten, Advances in atomic, molecular, and optical physics {\bf 37}, 181 (1996).

\bibitem{avalanches} J. Schuster, A. Marte, S. Amtage, B. Sang, G. Rempe, and H.C.W. Beijerinck,
Phys. Rev. Lett {\bf 87}, 170404 (2001).

\bibitem{verenna} W. Ketterle, D.S. Durfee, and D.M. Stamper-Kurn, \textit{Making, probing and understanding Bose-Einstein condensates}, Proceedings of the International School of Physics 'Enrico Fermi', Course CXI. IOS Press Ohmsha (1999).

\end{thebibliography}
\end{document}